\documentclass[structabstract]{aa}  
\usepackage{graphicx}
\usepackage{txfonts}
\newcommand{\Lsol}{L$_{\odot}$}
\newcommand{\Msol}{M$_{\odot}$}
\newcommand{\Msold}{M$_{\odot}$\,yr$^{-1}$}
\newcommand{\Vlsr}{V$_{\rm lsr}$}
\newcommand{\Vexp}{V$_{\rm exp}$}
\newcommand{\Vc}{V$_{\rm c}$}
\newcommand{\kms}{km\,s$^{-1}$}
\newcommand{\HI}{H\,{\sc {i}}~}
\newcommand{\MHI}{M$_{\rm H \sc I}$}
\newcommand{\NH}{N$_{\rm H}$}
\newcommand{\Av}{A$_{\rm v}$}
\newcommand{\Ks}{K$_{\rm s}$}
\newcommand{\cmd}{cm$^{-2}$}
\newcommand{\vsh}{\hspace*{0.1cm}}
\newcommand{\lsim}{~\rlap{$<$}{\lower 1.0ex\hbox{$\sim$}}}
\newcommand{\gsim}{~\rlap{$>$}{\lower 1.0ex\hbox{$\sim$}}}
\begin{document}
   \title{CO and \HI observations of an enigmatic interstellar cloud}

   \author{Y. Libert\inst{1},  
	   E. G\'erard\inst{2},
	   T. Le\,Bertre\inst{1},
           L.D. Matthews\inst{3},
	   C. Thum\inst{4},
	   \and J.M. Winters\inst{4}
          }

   \institute{LERMA, UMR 8112, Observatoire de Paris, 
	      61 Av. de l'Observatoire, 75014 Paris, France 
         \and
	      GEPI, UMR 8111, Observatoire de Paris, 
	      5 Place J. Janssen, 92195 Meudon Cedex, France
         \and
              MIT Haystack Observatory, Off Route 40,
              Westford, MA 01886, USA
         \and
              IRAM, 300 rue de la Piscine, 
	      38406 St. Martin d'H\`eres, France
             }

   \date{Received March 6, 2009; accepted April 6, 2009}

   \titlerunning{CO and \HI observations of an enigmatic cloud}
   \authorrunning{Libert et al.}

 
  \abstract
 {An isolated \HI cloud with peculiar properties has recently been discovered 
by Dedes, Dedes, \& Kalberla (2008, A\&A, 491, L45) 
with the 300-m Arecibo telescope, and subsequently imaged with the VLA. 
It has an angular size of $\sim$\,6$'$, and the \HI emission has 
a narrow line profile of width $\sim$ 3\,\kms.}
   {We explore the possibility that this cloud could be associated with a 
circumstellar envelope ejected by an evolved star.}
   {Observations were made in the rotational lines of CO with 
the IRAM-30m telescope, on three positions in the cloud, and a total-power 
mapping in the \HI line was obtained with the Nan\c cay Radio Telescope.}
   {CO was not detected and seems too underabundant in this cloud 
to be a classical late-type star circumstellar envelope. On the other hand, 
the \HI emission is compatible with the detached-shell model that we  
developed for representing the external environments of AGB stars.}
  {We propose that this cloud could be a fossil circumstellar shell left over 
from a system that is now in a post-planetary-nebula phase. 
Nevertheless, we cannot rule out that it is a Galactic cloud or 
a member of the Local Group, although the narrow line profile 
would be atypical in both cases.}


   \maketitle
%

\section{Introduction}

In the course of a sky survey for \HI halo clouds with the Arecibo 300-m 
Radio Telescope, Dedes et al. (2008, DDK2008) detected a cloud with 
unusual properties. It is isolated, nearly circular with an angular size of 
6.4$'$. The \HI emission is centered at \Vlsr = 47.6 \kms  ~and 
shows a surprisingly small linewidth of 3.4 \kms. It has a peak hydrogen 
column density, \NH = 60 10$^{18}$ \cmd. 
Follow-up observations with the Very Large Array (VLA) in the D-configuration 
show a slightly elongated structure, oriented at PA $\sim$ --14$^{\circ}$, 
and resolved into an elliptical ring of \HI emission peaks 
(from 1 to 1.8 10$^{20}$ \cmd), of about 4$'\times3'$, with the major axis 
along the same direction (cf. their figure~1b). 
A velocity gradient of $\sim$ 1 \kms ~is also seen along the major axis 
(their figure~1c). A faint star (2MASS\,07495348+0430238) was found coincident 
with the density minimum at the center of the ring. The difference between 
the images obtained at Arecibo and with the VLA is probably an effect of
the interferometric mode of observation which tends to filter extended 
emission (de Pater et al. 1991).

Several hypotheses are considered by DDK2008. Among them, that this enigmatic 
cloud, hereafter refered to as DDK cloud, is a circumstellar envelope around 
an evolved star. Indeed, the narrow line profile is typical of what is 
observed in \HI around this type of source (G\'erard \& Le\,Bertre 2006, 
Matthews \& Reid 2007). In addition the image obtained at the VLA is 
reminiscent of the fragmented ring, reported also in \HI by 
Rodr{\'\i}guez et al. (2002), in the Planetary Nebula NGC~7293 (Helix Nebula). 
With this line of thought, we have obtained new data in the CO rotational 
lines with the IRAM 30-m telescope, and in the \HI emission line at 21 cm with 
the Nan\c cay Radio Telescope. In this paper, we present our results and 
discuss in more detail the possibility that the DDK cloud is associated with 
the mass loss of an evolved star.


\section{Observations}

\subsection{\HI observations}\label{HI}

New \HI data have been obtained with the Nan\c cay Radio Telescope (NRT). 
This meridian telescope has a clear rectangular aperture with effective 
dimensions 160\,m$\times$30\,m. Thus the beam has a FWHM (full-width at 
half-maximum) of 4$'$ in right ascension (RA) and 22$'$ in declination (Dec) 
at 21 cm. The spatial resolution in RA is comparable to that 
of Arecibo. The point source efficiency is 1.4\,K\,Jy$^{-1}$, and the 
beam efficiency, 0.65.

\begin{figure}
\includegraphics[width=5.6cm,angle=-90]{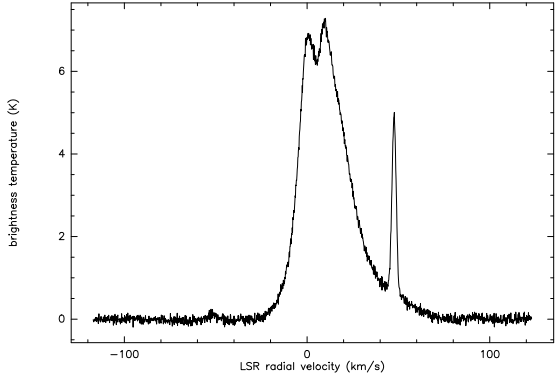}
\caption{Frequency-switch spectrum obtained with the NRT.}
\label{fswitch}
\end{figure}

A frequency-switch spectrum at a resolution of 0.16 \kms ~was first obtained 
on the source, defined by the 2MASS star. Galactic \HI 
emission is detected from --40 to 80\,\kms ~with a maximum of 7.3\,K at 
\Vlsr\,=\,9\,\kms ~(Fig.~\ref{fswitch}). The DDK cloud emission is clearly 
detected at \Vlsr\,=\,+47.6\,\kms ~over Galactic emission at a level of 
0.6\,K. The integrated Galactic emission in the DDK cloud line-of-sight is 
$\sim$\,265\,K\,\kms ~which translates to a hydrogen column density, 
\NH\,=\,4.8\,10$^{20}$ \cmd. Using the standard relation, 
\NH/\Av\,=\,1.87\,10$^{21}$ \cmd\,mag$^{-1}$, we derive a Galactic 
extinction over the line of sight, \Av\,=\,0.26. This estimate brings an 
upper limit to the extinction towards the DDK cloud, if it is located within 
our Galaxy. 

An additional grid of positions was mapped using the NRT. 
The observations have been obtained in the position-switch mode 
at a spectral resolution of 0.32 \kms,  
with again the on-position on the 2MASS star, and the off-positions at 
$\pm$\,2$'$, $\pm$\,4$'$, $\pm$\,6$'$, $\pm$\,8$'$, and $\pm$\,12$'$ in 
the east-west direction. The analysis has been performed as for the study 
of EP\,Aqr and Y\,CVn (Le\,Bertre \& G\'erard, 2004). In the direction of 
the DDK cloud, we encounter no effective confusion by Galactic \HI 
at \Vlsr$>$\,30\,\kms, in agreement with the inspection of the LAB Survey 
of Galactic \HI (Kalberla et al. 2005). 
We find no difference between the position-switch spectra at $\pm$\,8$'$ and 
at $\pm$\,12$'$, so that we can set an upper limit of 12$'$ for the DDK cloud 
extension in RA. It is noteworthy that these two position-switch spectra 
perfectly agree with the baseline-subtracted frequency-switch spectrum 
obtained with the telescope pointing directly on the 2MASS star (cf. 
Fig.~\ref{fswitch}, with a conversion factor, 2.15\,K/Jy).

\begin{figure}
\includegraphics[width=6.4cm,angle=-90]{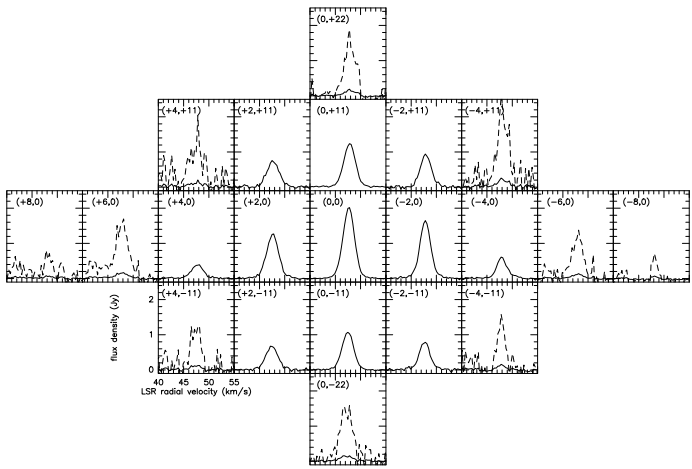}
\caption{Map of the 21 cm \HI emission of the DDK cloud 
obtained with the NRT. The central position 
corresponds to 2MASS\,07495348+0430238. The steps are 2$'$ in RA and 11$'$ in 
Dec; north is up, and east to the left. For the extreme positions, the 
spectra scaled by a factor 10 are also shown in dashed lines.}
\label{HImap}
\end{figure}

\begin{figure}
\includegraphics[width=5.6cm,angle=-90]{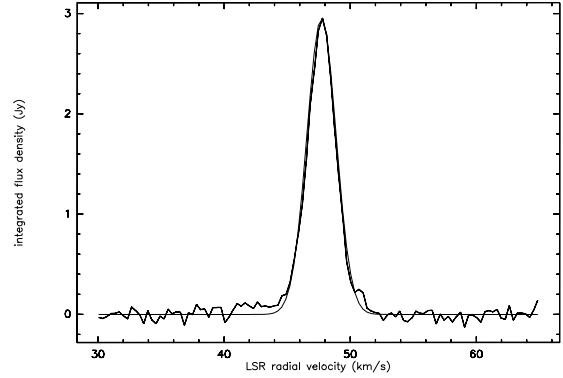}
\caption{Integrated \HI spectrum of the DDK cloud obtained with the NRT. 
A fit with one Gaussian (V$_{\rm c}$ = 47.74 \kms, FWHM = 2.73 \kms) 
is represented by a thin line. An excess of emission can be seen in 
the wings of the profile.}
\label{HIspectrum}
\end{figure}

We have also obtained data in the position-switch mode at +11$'$ (north) and 
--11$'$ (south) with off-positions at $\pm$\,2$'$, $\pm$\,4$'$,  
and $\pm$\,12$'$ (east-west), and at +22$'$ (north) and --22$'$ (south), 
with off-positions at $\pm$\,12$'$ (east-west). 
All these data are used to construct the \HI map that is 
presented in Fig.~\ref{HImap}. By integrating the individual spectra over 
the map, we obtain the integral spectrum of the DDK cloud that is presented in 
Fig.~\ref{HIspectrum}. 

\begin{table}
\centering
\caption{\HI line profile parameters obtained with the NRT data. 
In parentheses, we give formal errors resulting from the H\,{\sc {i}}-line 
Gaussian-fits (Landman et al. 1982).}
\begin{tabular}{ccccc}
\hline
         & \Vc           &     FWHM      & Intensity    & Integrated flux\\
         & (\kms)        &    (\kms)     & (Jy)         & (Jy$\times$\kms)\\
\hline
 4$'$ W  & 47.96\,(0.02) &  2.56\,(0.03) & 0.59\,(0.01) & 1.62\,(0.05)\\
 2$'$ W  & 47.80\,(0.01) &  2.61\,(0.01) & 1.65\,(0.01) & 4.58\,(0.05)\\
 ''on''     & 47.64\,(0.01) &  2.72\,(0.01) & 2.02\,(0.01) & 5.84\,(0.01)\\
 2$'$ E  & 47.58\,(0.01) &  2.91\,(0.02) & 1.24\,(0.01) & 3.85\,(0.05)\\
 4$'$ E  & 47.70\,(0.04) &  2.94\,(0.06) & 0.39\,(0.02) & 1.23\,(0.05)\\
\hline
22$'$ N & 47.89\,(0.10) &  2.44\,(0.10) & 0.15\,(0.01) & 0.38\,(0.05)\\
11$'$ N & 47.83\,(0.01) &  2.82\,(0.01) & 1.24\,(0.01) & 3.72\,(0.02)\\
11$'$ S & 47.48\,(0.01) &  2.75\,(0.01) & 1.06\,(0.01) & 3.12\,(0.02)\\
22$'$ S & 47.16\,(0.10) &  3.45\,(0.15) & 0.15\,(0.01) & 0.56\,(0.05)\\
\hline
Total       & 47.74\,(0.01) &  2.73\,(0.02) & 2.95\,(0.03) & 8.59\,(0.11)\\
\hline
\end{tabular}
\label{gauss}
\end{table}

This spectrum shows a Gaussian-like profile of width, FWHM = 2.73\,\kms, 
and centered at \Vlsr = 47.74\,\kms 
~(Table~\ref{gauss}, ``Total''). The \HI line profile is narrow and 
can be used to set an upper limit on the average hydrogen kinetic temperature 
of $\sim$~170~K. Emission in excess of the Gaussian profile may be present 
from 42 to 53 \kms ~at a level of $\sim$~20\,mJy. The integrated area is 
8.6\,Jy$\times$\kms, which translates to a hydrogen mass 
for the enigmatic cloud, 
at a distance $d$ expressed in kpc, \MHI\,=\,2.03$\times$\,$d^2$\,\Msol. 
The line profiles at the other positions in the map are also Gaussian-like, 
the intensity being slightly larger west than east. The center of mass 
of the \HI emission thus appears offset by --0.3\,$\pm$\,0.1$'$ (west) 
with respect to the 2MASS star. The source is slightly resolved in RA,  
with a size of 4.3\,$\pm$\,0.3$'$ (FWHM). 
There is also a possible offset in Dec., +1.1\,$\pm$\,0.3$'$ (north), 
and an extension that we estimate at $\sim$ 9\,$\pm$\,3$'$. 
Finally the centroid velocity is redshifted north and west, and blueshifted 
south and east as compared to the center (see Table~\ref{gauss}), 
in agreement with the velocity gradient reported by DDK2008. 

In general, our results agree with those of DDK2008. We confirm that 
the source is isolated and compact, and that it shows an ordered velocity 
gradient. However, we find a narrower profile (FWHM = 2.73\,\kms, 
Table~\ref{gauss}), than theirs (3.4\,$\pm$\,0.18\,\kms). 
Our estimate is also consistent with that obtained independently on the 
baseline subtracted frequency-switch spectrum presented in Fig.~\ref{fswitch} 
(FWHM = 2.91\,$\pm$\,0.1\,\kms).

\subsection{CO observations}\label{CO}

The DDK \HI cloud was observed in CO (1-0) and (2-1) with the IRAM-30m 
telescope on Dec. 5, 2008. We selected three positions, centered on peaks of 
\HI emission visible on the VLA map (Table~\ref{pos}), because in the Helix 
Nebula \HI emission was found in the outer parts where CO is 
also present (Rodr{\'\i}guez et al. 2002, Young et al. 1999). 
The telescope beamwidths (FWHM) are 21$''$ at 115 GHz and 11$''$ at 230 GHz, 
and are thus smaller than the VLA synthesized beam (45$''\times$35$''$). 
The data were obtained with the VESPA autocorrelator at different
spectral resolutions and bandwidths ~(resolution 10\,kHz and 20\,kHz,
bandwidths 35\,MHz and 53\,MHz, respectively at 3\,mm and resolution 20\,kHz 
and 40\,kHz with bandwidths of 35\,MHz and 107\,MHz, 
respectively, at 1\,mm) and, simultaneously, with a low resolution
filter bank (resolution 1\,MHz, bandwidth 256\,MHz). 
The system temperature was 450\,K at 3\,mm, and 700\,K at 1\,mm. 
We obtained spectra with an rms noise of 0.016 K (T$_{\rm mb}$) at 115 GHz, 
for a resolution of 2.6 \kms, and of 0.063 K at 230 GHz, for a resolution 
of 1.3 \kms. No emission was detected in any of the three positions.

We follow the Jura et al. (1997) approach for estimating upper limits 
on the CO column densities in the three lines-of-sight. 
We assume that CO is optically thin and warm ($\gg$\,11\,K), and 
use their equation (4):\\
N(CO) = 4.32 ~10$^{13}$ ~T$_{\rm ex}$ ~$\int$ T$_{\rm mb}$ dV, 
\vsh for CO(1-0), and \\
N(CO) = 1.08 ~10$^{13}$ ~T$_{\rm ex}$ ~$\int$ T$_{\rm mb}$ dV, 
\vsh for CO(2-1),\\
with N(CO) in cm$^{-2}$ and V in \kms. For T$_{\rm ex}$, we adopt, as an upper 
limit, the upper limit on the \HI kinetic temperature that was obtained in 
the previous section (T$_{\rm K}$ = 170\,K). By integrating our CO spectra 
over the range 42--53 \kms, the maximum velocity range over which we 
found \HI emission (cf. Sect.~\ref{HI}), we can thus derive 
conservative upper-limits of 6\,10$^{14}$ \cmd ~and 4\,10$^{14}$ \cmd, 
respectively from the two lines, on the column density in CO.
This may be compared to the peak CO column density of 
$\sim$\,1.5\,10$^{16}$ \cmd ~obtained by Young et al. (1999) at an angular 
resolution of 31$''$ in the Helix Nebula, for a corresponding peak \HI density 
of 1.2\,10$^{20}$ \cmd ~(Rodr{\'\i}guez et al. 2002).

\begin{table}
\centering
\caption{Positions observed in CO with the IRAM 30-m telescope.}
\begin{tabular}{rllcc}
\hline
       & $\alpha$ (2000.0) & $\delta$ (2000.0) & l$^{\rm II}$ & b$^{\rm II}$\\
\hline
\HI peak: A1 &  07 49 54    &  04 32 30   & 215.555 & 15.068\\
          A2 &  07 49 50    &  04 32 20   & 215.550 & 15.052\\
          A3 &  07 49 54    &  04 29 00   & 215.609 & 15.042\\
\hline
\end{tabular}
\label{pos}
\end{table}


\section{Discussion}

The DDK \HI cloud has no counterpart at other wavelengths. It is not seen 
on the IRAS maps and it has not been detected at 870 $\mu$m with the Large 
Bolometer Camera on the 12-m APEX antenna (DDK2008), showing no detection 
of cold dust, and no evidence of heating by an internal source. There is 
no obvious counterpart on the NRAO VLA Sky Survey (NVSS; Condon et al. 1998) 
continuum map obtained at 1.4 GHz. 

However, a 2MASS star was pointed out by DDK2008 at about the center of the 
cloud. The cross-identification with the USNO-B1.0 and Tycho-2 catalogs 
shows this star has an apparent proper motion of +4 mas/yr in RA and 
+30 mas/yr in Dec (PA = --8$^{\circ}$). The distance is not known. 
Its near-infrared colors (Table~\ref{Phot}) correspond to that of 
an F-type star, but the optical data correspond rather to an A-type star. 
It has no mid or far-infrared counterpart, but has been detected 
in the ultraviolet by GALEX.

There are also several faint stars around the DDK cloud. Interestingly among 
them there is a high proper-motion one at about 8$'$ south moving away 
from the \HI cloud (NLTT\,18499, +56 mas/yr in RA and --190 mas/yr 
in Dec, Lepine \& Shara 2005). 

\begin{table}
\centering
\caption{Photometry of 2MASS\,07495348+0430238 and NLTT\,18499, without  
correction, and with a correction corresponding to \Av = 0.26 (R=3.1; 
IR/Visual\,: Fitzpatrick 1999, UV\,: Rey et al. 2007). Sources: 
IR (2MASS), Visual (Tycho-2, USNO-B1.0, Droedge et al. 2006), UV (GALEX, 
AB system).}
\begin{tabular}{llllll}
\hline
     & 2MASS   &             &  &  NLTT   & \\
     & \Av = 0 & \Av = 0.26  &  & \Av = 0 & \Av = 0.26 \\
\hline
\Ks  & 10.22   & 10.19       &  & 13.58   & 13.55 \\
H    & 10.24   & 10.19       &  & 13.74   & 13.69 \\ 
J    & 10.50   & 10.43       &  & 14.24   & 14.07 \\
I    & 10.94   & 10.81       &  & 15.1    & 15.0  \\
R    & 11.6    & 11.4        &  & 16.0    & 15.8  \\
V    & 11.6    & 11.4        &  & --      & --    \\
B    & 11.7    & 11.4        &  & 17.3    & 17.0  \\
NUV  & 15.57   & 14.82       &  & --      & --    \\
FUV  & 21.4    & 20.7        &  & --      & --    \\
\hline
\end{tabular}
\label{Phot}
\end{table}

\subsection{The circumstellar-shell hypothesis}

Following DDK2008, we adopt arbitrarily a distance of 400 pc, which translates 
to a height above the Galactic Plane of 100 pc (b$^{\rm II}$ = +15$^{\circ}$). 
The DDK cloud mass in atomic hydrogen would therefore be 0.32 \Msol, 
and the size, 0.6 pc. These estimates are typical of circumstellar shells 
around evolved red giants, carbon stars or Planetary Nebulae (G\'erard 
\& Le\,Bertre 2006). In this context, as the \HI mass scales as $d^2$, 
the DDK cloud cannot be much further than 1 kpc. The narrow and Gaussian-like 
line profile is also typical of those obtained for such sources. Libert 
et al. (2007) have developed a model in which such a line profile results 
from the slowing-down of a stellar wind by ambient matter. In this model a 
``detached shell'' is built over time, with an inner radius where the stellar 
outflow is abruptly slowed-down (termination shock) and an outer radius 
where external matter is compressed by the expanding shell (bow shock). 
The detached shell is thus formed of compressed circumstellar and interstellar 
materials, which are heated when crossing the shocks, and cooling-down 
after that. They applied this model to the detached shell observed around 
the carbon star Y CVn (Izumiura et al. 1996), and were able to reproduce 
satisfactorily the \HI line profiles obtained at different positions.
 
\begin{figure}
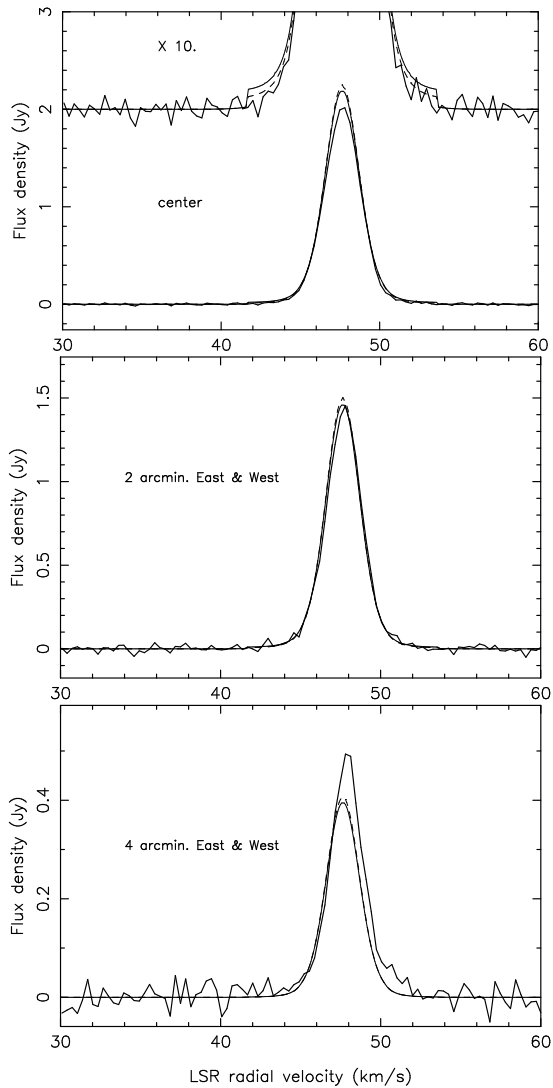

\centering
\includegraphics[width=4.6cm,angle=-90]{DDK-C.ps}
\includegraphics[width=4.6cm,angle=-90]{DDK-0p5EW.ps}
\includegraphics[width=5.1cm,angle=-90]{DDK-1EW.ps}
\caption{Comparison between the \HI line profiles obtained with the NRT 
(thick lines) and the model of a detached shell around an AGB star discussed 
in Sect. 3.1 (thin and dashed lines). Top: 
central spectrum; middle: average of the two spectra at +2$'$ (east) 
and --2$'$ (west); bottom: average of the two spectra at +4$'$ and --4$'$.}
\label{model}
\end{figure}

\begin{figure}
\centering
\includegraphics[width=5.cm,angle=-90]{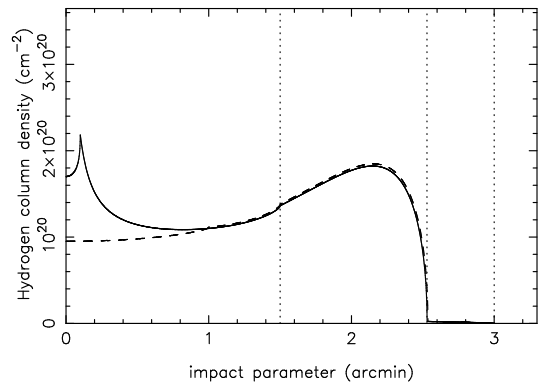}
\caption{Atomic hydrogen column density profile for the detached shell 
model discussed in Sect.~3.1. The vertical dotted lines mark the radii 
r$_1$, r$_f$ and r$_2$ of the model (see Table~\ref{modelfit}).}
\label{columndensity}
\end{figure}

In order to check the applicability of this model to the DDK cloud, 
we have performed a calculation with the parameters given in Table~4. 
We take a star undergoing mass loss,  
at a rate of 3\,10$^{-7}$\,\Msold ~(in atomic hydrogen) and with an expansion 
velocity, \Vexp = 6 \kms, for $\sim$ 10$^6$ years. The expansion velocity has 
been selected such as to cover the range of \HI emission, from 42 to 53 \kms. 
In Fig.~\ref{model}, 
the results (solid lines) are compared to observations. 
As the model is spherical, we have averaged the east and west spectra.
The \HI line profiles and intensities are reproduced satisfactorily. 
In Fig.~\ref{columndensity}, we present the \HI column density derived 
from the model; it gives a peak \NH $\sim$ 1.8 10$^{20}$ \cmd ~at 2.1$'$ 
in accordance with the VLA results (DDK2008). However, in contrast to the 
Y CVn detached shell, there is no evidence of a red giant associated to 
the DDK \HI cloud (see below). Therefore the model that is presented here 
only demonstrates that the observed properties of the enigmatic cloud, 
and in particular its peculiar \HI line profile, 
can be easily accounted for by mass loss from a stellar source. 

\begin{table}
\centering
\caption{Model parameters (d = 400 pc). The notations are the same 
as in Libert et al. (2007).}
\begin{tabular}{ll}
\hline
\.M (in hydrogen)                 & 3 10$^{-7}$ \Msold\\
$\mu$                             & 1.3\\
t$_1$                             & 5.7 10$^{4}$ years\\
t$_{DS}$                          & 9.7 10$^{5}$ years\\
r$_1$                             & 0.17 pc (1.5$'$)\\
r$_f$                             & 0.29 pc (2.53$'$)\\
r$_2$                             & 0.35 pc (3$'$)\\
T$_0$($\equiv$ T$_1^-$), T$_1^+$  & 20 K, 1070 K\\
T$_f$ (= T$_2$)                   & 92 K\\
v$_0$($\equiv$ v$_1^-$), v$_1^+$  & 6 \kms, 1.5 \kms\\
v$_f$                             & 0.04 \kms\\
v$_2$                             & 0.6 \kms\\
n$_1^-$, n$_1^+$                  & 5.2 H\,cm$^{-3}$, 21.0 H\,cm$^{-3}$\\
n$_f^-$, n$_f^+$                  & 301.0 H\,cm$^{-3}$, 2.2 H\,cm$^{-3}$\\
n$_2$                             & 1.6 H\,cm$^{-3}$\\
M$_{r < r_1}$ (in hydrogen)       & 1.7 10$^{-2}$ \Msol\\
M$_{DT,CS}$   (in hydrogen)       & 0.29 \Msol\\
M$_{DT,EX}$   (in hydrogen)       & 3.3 10$^{-3}$ \Msol\\
\hline
\end{tabular}
\label{modelfit}
\end{table}

The \HI image of the DDK cloud obtained with the VLA by DDK2008 
shows some similarities to that observed by Rodr{\'\i}guez et al. (2002) 
in the Helix Nebula (NGC\,7293), a Planetary Nebula (PN) and therefore 
a source in a late stage of evolution after the Asymptotic Giant 
Branch (AGB). Their data obtained with the VLA in the DnC configuration reveal 
a ring of atomic hydrogen, with the \HI emission concentrated in clumps. 
This \HI ring has a diameter of 12$'$, or 0.7 pc at a distance of 200 pc.
The emission coincides with the continuum emission 
at 1.4 GHz and seems to delineate the outer parts of the ionized gas.  
G\'erard \& Le\,Bertre (2006) 
estimate the total atomic hydrogen mass in the Helix Nebula at 0.26 \Msol. 
CO emission has also been detected around the ionized gas (Young et al. 1999). 
It delineates the same ring as in H\,{\sc {i}}, but is also found in small 
cometary globules embedded in the ionized gas (Huggins et al. 2002).

On the other hand, there are important differences between 
the Helix Nebula and the DDK cloud. The Helix Nebula exhibits 
a much broader global line profile in \HI (FWHM $\sim$ 35 \kms, 
G\'erard \& Le\,Bertre 2006). Molecular gas (CO, Sect. 2.2) has not been 
detected in the DDK cloud at the three peaks of \HI column 
density that have been observed, which weakens our working hypothesis unless 
CO is concentrated in small globules that we have missed. The Helix Nebula, 
like many other PNs, has been detected in continuum emission at 1.4 GHz 
in the NVSS, whereas no such emission is detected for the DDK cloud. IRAS 
and ISO observations of the Helix Nebula at 90 and 160 $\mu$m show extended 
($\sim 20 '$) 
thermal emission by dust with a possible 
contribution of emission lines (Speck et al. 2002). There is no evidence for 
thermal emission by dust in the DDK cloud (DDK2008).
The Helix Nebula is a well-known emission-line source (PK 036-57), but there 
is no emission-line source associated with the DDK cloud. Also an  inspection 
of the Southern H$\alpha$ Sky Survey Atlas (Gaustad et al. 2001) shows 
no emission close to its position.  
Thus there is  presently no evidence for ionized material 
in the DDK cloud, and it seems difficult to associate it with a PN. 

DDK2008 suggest that the \HI cloud could be formed by mass loss from the star 
detected in the 2MASS survey. 
The colors of this star are not compatible with those of red giants. 
At a distance of 400 pc, its luminosity should be on the order of 3$\pm$1 
\Lsol ~(depending on its exact energy distribution), much too low for being 
a star in transition between the AGB and the PN stages. In fact from the 
non-detections at IRAS wavelengths and at 870 $\mu$m, there is no evidence 
of a luminous star inside the DDK cloud. 

{\it In the context of the circumstellar-shell hypothesis}, there is only 
one option left: that the DDK cloud is a fossil circumstellar shell of 
a source that is now in a post-PN phase. This source would then be evolving 
towards the white dwarf stage, i.e. it would be a 
stellar core of decreasing luminosity and temperature such that 
there is no significant heating nor ionization close to the DDK cloud. 
An obvious candidate would be 2MASS\,07495348+0430238. 
However this object seems peculiar: the optical and near-infrared colors do 
not match well. It could be a variable star, a star with an infrared excess, 
or a binary system.
Also, the stellar remnant might have moved away 
from its circumstellar shell (e.g. Smith 1976), and nearby stars, 
such as NLTT\,18499, should also be considered. 
Radial velocity measurements could help to select the best candidate. 
A parallax would constrain the distance, and, {\it in this context}, 
the physical characteristics of the DDK cloud. 

Due to the large scale of the circumstellar shell, a cessation of the mass 
loss by the central star would not affect strongly the \HI emission before 
several 10$^4$ years (i.e. before a lapse smaller than the time needed 
by stellar matter to reach the termination shock in r$_1$, 5.7 10$^4$ years). 
In order to illustrate this effect, we have made a second run of our model, 
with only t$_1$ reduced from 
5.7 10$^4$ years to 2 10$^4$ years, all other input parameters being kept 
equal. Only the central part (r\,$<$\,r$_1$) is affected by this modification 
because the wind is supersonic up to the termination shock. The results 
are shown in dashed lines in Figs.~\ref{modelfit} and \ref{columndensity}. 
The predicted {H\,{\sc {i}}}-profiles are almost indistinguishable, 
except in the wings of the central-position line profile.

We should also account for the lack of H$\alpha$ and CO emission. 
For an average density of 10 cm$^{-3}$, the recombination time of electrons 
is $\sim$ 10$^{4}$ years, which sets a lower limit on the time since 
the hypothetical source of ionization should have been switched-off. 
The timescale for CO 
photo-dissociation in the ISM is $\sim$ 200 years (Mamon et al. 1988). 
However, CO is self-shielded and may survive for a much longer time, 
up to 10$^{5}$ years, depending on the mass-loss rate and expansion velocity, 
and possibly more if the medium is inhomogeneous.  
The timescale for the dispersion of circumstellar shells around evolved stars 
is also uncertain. It probably depends on the wind history, on the properties 
of the ambient ISM and on the velocity of the central star relative to 
this medium (Villaver et al. 2002, 2003). \HI observations show that 
these structures may have a lifetime of at least 
several 10$^5$ years (G\'erard \& Le\,Bertre 2006, Libert et al. 2007). 
Some post-PN circumstellar shells, old enough to escape 
detection in H$\alpha$ and CO, should then be expected. 
However, up to now none has been identified. The DDK cloud would then be 
the first specimen, and we probably need to identify other cases before 
reaching a firm conclusion on the circumstellar-shell hypothesis.

\subsection{The IVC/HVC hypothesis}

The DDK cloud could also be associated with the population of high galactic 
latitude clouds that are found at high velocity, $|$\Vlsr$|$ $>$ 70 \kms 
~(High-Velocity Clouds, HVC), or at lower velocity 
(Intermediate-Velocity Clouds, IVC).
In particular de Heij et al. (2002a) have identified a population of small 
($\leq$ 1$^{\circ}$ FWHM) and isolated high-velocity clouds (compact HVC, 
or CHVC) sharply bounded in angular extent. A follow-up study with the 
Westerbork Synthesis Radio Telescope shows a core-halo morphology similar 
to that seen in the DDK \HI cloud (de Heij et al. 2002b, 
see e.g. CHVC 120-20-443 in their figure 2). 
The enigmatic cloud has a radial velocity lower than the characteristic 
velocity of HVC, but we have no information on its transverse velocity. 
Perhaps more confounding, the velocity dispersion in CHVC (typically 20 \kms) 
is much larger than that in the DDK cloud. IVC have probably a two-phase 
structure with bright clumps and a diffuse envelope similar to that of HVC 
(Smoker et al. 2002, Haud 2008). 

Absorption H\,{\sc {i}}-line surveys made with Arecibo and the Giant 
Metrewave Radio Telescope (Heiles \& Troland 2003, Mohan et al. 2004) have 
revealed the presence in the ISM of many small clouds, at low and intermediate 
radial velocity, with hydrogen at a temperature in the range 50-200\,K, 
comparable to that in the DDK cloud. These clouds might be associated 
to the population of discrete \HI clouds discovered in emission 
with the Green Bank Telescope by Lockman (2002). The latter seem to follow 
Galactic rotation, to have a peak \NH ~of a few times 10$^{19}$ \cmd, and
linewidths in the range of a few to tens of \kms. 
Lockman (2002) proposes that they are located in the Galactic halo, 
and have sizes of few tens of parsecs and typical masses of 50 \Msol. 
We cannot exclude that the DDK cloud would be a member of this population, 
on the low side of its velocity-dispersion distribution. 

The origin of CHVC/IVC is a matter of debate, the most critical difficulty 
being their uncertain distance. In fact as, up to now, 
post-PN circumstellar shells have still not been identified, 
we would like to raise the possibility that some of these objects 
may hide among the population of CHVC/IVC. D\'esert et al. (1990) 
have already noted a coincidence between an IVC, which they detected in CO, 
and a white dwarf. It is also known that some evolved stars 
at high-galactic latitude are associated with 
extended gaseous tails that show a cometary morphology in \HI 
evocative of IVC/HVC (Matthews et al. 2008, Libert et al. 2008).

\subsection{The extragalactic hypothesis}

DDK2008 raised the possibility that the \HI cloud they discovered might be 
extragalactic in origin. In this case, the requirement that the cloud be 
gravitationally bound imposes an upper limit on its distance of $\sim$530~kpc 
(in order that its \HI mass does not exceed its virial mass). 
Placed at a nearer distance, the \HI mass alone would no longer 
be able to account for the observed 
linewidth, implying that some additional ``dark'' component must be present. 
DDK2008 therefore suggested that the cloud might be an example of a ``dark 
galaxy''---a galactic system too low in mass to have become unstable to star 
formation. This hypothesis is of particular interest, since the existence of 
large numbers of low-mass satellites to the Milky Way (the smallest of which 
are not expected to have formed stars) has been predicted by cold dark matter 
models of galaxy formation
(e.g., Klypin et al. 1999; Moore et al. 1999).

Recently Ryan-Weber et al. (2008) discovered an unusual dwarf galaxy, Leo~T, 
whose \HI properties share some interesting similarities with the DDK cloud. 
Both Leo~T and the DDK cloud have similar angular sizes ($\sim5'$) and clumpy, 
elliptical-shaped \HI morphologies. Placed at the adopted distance of Leo~T 
(420~kpc), the DDK cloud and the Leo~T dwarf would also have comparable 
\HI masses ($3.4\times10^{5}~M_{\odot}$ and $2.8\times10^{5}~M_{\odot}$, 
respectively). However, there are several noteworthy differences between these 
two objects. The Leo~T dwarf has a larger \HI velocity width ($\sim$7~\kms), 
and its velocity field shows no signatures of rotation. Indeed, lack of 
ordered rotation tends to be a generic feature of the lowest mass dwarf 
galaxies (e.g., Grebel 2008). Another key difference is that Leo~T has a ratio 
of dynamical mass to \HI mass $\sim$50 (and very few stars), implying a large 
dark matter fraction. In contrast, $M_{\rm dyn}/M_{\rm HI}\sim$2 for the DDK 
cloud if located at the same distance. After accounting for the mass 
contribution of helium, this leaves little room for a significant amount 
of ``dark'' material.

If the DDK cloud is truly a rotating disk, then its measured rotational 
velocity should be corrected for the disk's inclination to our line-of-sight. 
Assuming the disk is circular with an intrinsic flattening $q\sim$0.1, its 
measured \HI axial ratio ($b/a$=0.7; DDK2008) implies $i\approx46^{\circ}$, 
based on the standard relation \\
\hspace*{1.cm} ${\rm cos}^{2}i = \frac{\left(\frac{b}{a}\right)^{2} - 
q^{2}_{0}}{1 - q^{2}_{0}}$\\
Thus the true peak rotational velocity is 
$V_{\rm rot}/{\rm sin}i\approx$1.4~\kms. Yet, this value is still 
extraordinarily small, leading to a difficulty in explaining how such a 
low-mass system could have retained an observable quantity of cold, 
neutral gas to the present day.

Current galaxy formation models predict that the reionization of the 
intergalactic medium at high redshift will suppress gas accretion onto 
galactic potentials with circular velocities $V_{\rm circ}\lsim$20-30~\kms\ 
(e.g., Bullock et al. 2000). Even if such low-mass structures were able to 
collapse, their gas would be expected to rapidly photoevaporate (Barkana 
\& Loeb 1999). Furthermore, gas would be prevented from condensing back onto 
such structures during later epochs, since the characteristic mass scales for 
structure formation in the intergalactic medium exceed the mass of these 
``mini-halos'' (e.g., Gnedin 2000).

Ricotti (2009) proposed that this latter problem could be partly overcome by 
the increasing central concentration of the galaxy potentials and the 
decreasing temperature of the intergalactic medium as a function of decreasing 
redshift. However, his models predict that the lowest mass galaxies able to 
cool below $10^{4}$~K within a Hubble time have 
$V_{\rm circ}\sim$5-7~\kms---several times higher than the DDK cloud. 
Moreover, the dynamical masses of the smallest galaxies are predicted to 
exceed their gas masses by more than an order of magnitude. The latter 
discrepancy could be alleviated by placing the DDK cloud at a smaller distance 
(e.g., comparable to the Magellanic Clouds). However, in this case, any 
accreted gas would likely be depleted via ram pressure stripping during 
passage through the Galactic corona (Mayer et al. 2006).

Finally, there may be some difficulty accounting for the presence of an \HI 
column density minimum or ``hole'' near the center of rotation of the DDK 
cloud in an extragalactic scenario. While such features are common in dwarf 
galaxies, their origins are most likely tied either directly or indirectly 
with star formation, arising from energy injection from stellar winds and/or 
supernovae (e.g., Kerp et al. 2002), or from a combination of turbulence and 
thermal and gravitational instabilities (Dib \& Burkert 2005). In summary, 
while an extragalactic origin for the DDK cloud cannot yet be strictly 
excluded, it appears unlikely in light of the available data for the cloud 
and our present theoretical understanding of the formation and evolution of 
the lowest mass galaxies.

\section{Conclusions} 

The \HI emission from the DDK cloud is well separated from the rest of the 
Galactic emission. It shows a narrow line profile that can be fitted with 
one Gaussian of width 2.8\,\kms ~and that is centered at \Vlsr\,=\,47.7\,\kms.
The cloud has a size of 4$'$ in RA, and for a distance $d$ 
expressed in kpc, a mass in atomic hydrogen of 2$\times$$d^2$\,\Msol. 
The rotational lines of CO (2-1 and 1-0) have not been detected.

The \HI line profiles are compatible with the model of a detached shell 
around an AGB star, which we have developed for the carbon star Y CVn. 
However, owing to the absence of a luminous and/or hot 
central star, we discard the possibility that the DDK cloud is related to 
a mass-losing red giant, a post-AGB object or a planetary nebula central star. 
In the context of the circumstellar shell hypothesis, we suggest that 
the DDK cloud could be a fossil shell left over by a stellar core 
(still to be identified, but possibly associated with the 2MASS star 
pointed out by DDK2008) that is now evolving towards the white-dwarf stage. 

With a core-halo morphology, the DDK \HI cloud might also be related to a  
compact HVC/IVC, although the narrow \HI linewidth would be atypical. 
An extragalactic origin can also be considered, but again appears improbable 
in view of the small velocity dispersion. 

Presently the circumstellar shell hypothesis is the only one that can easily 
account for the small linewidth. If this hypothesis proves to be correct, 
the DDK cloud might offer the first occasion to study a post-PN stellar 
remnant together with its fossil shell.

\begin{acknowledgements}
The Nan\c{c}ay Radio Observatory is the Unit\'e scientifique de Nan\c{c}ay of 
the Observatoire de Paris, associated as Unit\'e de Service et de Recherche 
(USR) No. B704 to the French Centre National de la Recherche Scientifique 
(CNRS). The Nan\c{c}ay Observatory also gratefully acknowledges the financial 
support of the Conseil R\'egional de la R\'egion Centre in France.
IRAM is supported by INSU/CNRS (France), MPG (Germany), and IGN (Spain).
This research has made use of the SIMBAD and VizieR databases, operated 
at CDS, Strasbourg, France and of the NASA's Astrophysics Data System.
\end{acknowledgements}

\end{document}